\documentclass[preprint,times,12pt,nonatbib]{elsarticle}
\makeatletter
\let\c@author\relax
\makeatother

\usepackage{amsbsy}
\usepackage{amssymb}
\usepackage{amsmath}
\newcommand\aug{\fboxsep=-\fboxrule\!\!\!\fbox{\strut}\!\!\!}
\usepackage{placeins}
\usepackage{csquotes}
\usepackage{bm}
\usepackage{doi}

\usepackage[backend=bibtex,
            hyperref=true,
            style=authoryear,
            natbib=true,
            doi=true,
            url=false,
            isbn=false, 
            uniquename=false, 
            uniquelist=false]{biblatex}
\usepackage{hyperref}
\hypersetup{colorlinks}

\usepackage{lineno}

\journal{Computers, Environment and Urban Systems}

\begin{document}

\begin{frontmatter}



\title{PyLUSAT: An open-source Python toolkit for GIS-based land use
suitability analysis}


\author[aff1,aff2]{Changjie Chen}
\author[aff2]{Jasmeet Judge}
\author[aff1]{David Hulse}
\affiliation[aff1]{
            organization={Florida Institute for Built Environment Resilience,
            University of Florida},
            addressline={606 SE Depot Ave},
            city={Gainesville},
            postcode={32611}, 
            state={FL},
            country={United States}
            }
\affiliation[aff2]{
            organization={Center for Remote Sensing,
            Department of Agricultural and Biological Engineering,
            University of Florida},
            addressline={1741 Museum Road},
            city={Gainesville},
            postcode={32611}, 
            state={FL},
            country={United States}
            }

\begin{abstract}
Desktop GIS applications, such as \textit{ArcGIS} and \textit{QGIS}, provide
tools essential for conducting suitability analysis, an activity that is
central in formulating a land-use plan.
But, when it comes to building complicated land-use suitability models, these
applications have several limitations, including operating system-dependence,
lack of dedicated modules, insufficient reproducibility, and difficult, if not
impossible, deployment on a computing cluster.
To address the challenges, this paper introduces PyLUSAT: Python for Land Use
Suitability Analysis Tools.
PyLUSAT is an open-source software package that provides a series of tools
(functions) to conduct various tasks in a suitability modeling workflow.
These tools were evaluated against comparable tools in \textit{ArcMap 10.4}
with respect to both accuracy and computational efficiency.
Results showed that PyLUSAT functions were two to ten times more efficient
depending on the job's complexity, while generating outputs with similar
accuracy compared to the ArcMap tools.
PyLUSAT also features extensibility and cross-platform compatibility.
It has been used to develop fourteen QGIS \textit{Processing Algorithms} and
implemented on a high-performance computational cluster (H  iPerGator at the
University of Florida) to expedite the process of suitability analysis.
All these properties make PyLUSAT a competitive alternative solution for urban
planners/researchers to customize and automate suitability analysis as well as
integrate the technique into a larger analytical framework.
\vspace{12pt}
\end{abstract}

\begin{highlights}
\item {PyLUSAT is an open-source Python package dedicated to land-use
suitability analysis}
\item {An alternative solution for urban planners to conduct GIS-based
suitability modeling}
\item {Enabling rapid, massive simulation of future land-use scenarios in
high-performance computing clusters}
\item {PyLUSAT functions are computationally efficient, highly extensible, and
cross-platform compatible}
\item {It has been successfully implemented on HiPerGator, a supercomputer at
the University of Florida}
\end{highlights}


\end{frontmatter}


\section{Introduction}
Since first being introduced in late 1960s, Geographic Information System
(GIS)-based suitability analysis continues playing a significant role in
contemporary practices of land use planning \citep{Collins2001}.
It is particularly favored by land-use practitioners because the technique can
effectively synthesize spatial analytics, expert knowledge, and community
values, all of which, from a planning perspective, are critical factors to
consider when making land-use decisions.
In practice, suitability analysis is usually performed using GIS applications
with a Graphical User Interface (GUI), including ArcGIS and QGIS
\citep{Abdullahi2015,Mesgaran2017}.
And since both software applications provide a \textit{Python} package (namely
ArcPy and PyQGIS) as an interface to their respective functions, suitability
analysis can also be performed through scripting.
Because it allows researchers to customize and automate the analysis, the
second approach is preferable when dealing with complex processes. 

But tools offered by GIS applications are either constrained to run only
within their applications or they require the presence of the main application.
Additionally, some GIS application, like ArcGIS, are proprietary and Operating
System (OS)-dependent.
These restrictions of GIS applications act as a barrier to the integration of
suitability analysis into a more comprehensive framework that involves a series
of analytical components completed by multiple software.
Instead, a programming library is an alternative solution which nullifies such
integration barrier.
At the time of writing, there exist a wide variety of programming libraries
that are developed to study urban-related issues, with examples being
\textit{UrbanSim}—simulating urban real estate markets,
\textit{OSMnx}—analyzing street networks, and
\textit{UrbanAccess}—measuring transit accessibility
\citep{Waddell2010, Boeing2017, Blanchard2017}.

However, despite being an influential technique in land use planning, there
lacks a programming library that is dedicated to suitability analysis.
To fill in this gap, we present a free and open-source software (FOSS)
package—PyLUSAT: Python for Land-Use Suitability Analysis Tools.
PyLUSAT provides a complete set of tools that is capable of carrying out
geospatial analyses entailed by suitability analysis; serves as an alternative
solution for planners and researchers to perform suitability analysis; and
offers a cross-platform programming library that can run on a
\textit{supercomputer} (most of which are based on Linux OS) to take advantage
of High-performance computing (HPC) \citep{Strohmaier2020}.

In this paper, Section 2 reviews the literature and explains the choice of
adopting a vector-based GIS approach.
Section 3 describes the three major categories of PyLUSAT functions including
geospatial, transformation, and aggregation functions, which support the main
tasks in a suitability analysis.
Section 4 presents validation and performance evaluation of PyLUSAT
functions.
Finally, Section 5 summarizes the paper and briefly discusses future research.

\section{Literature Review}

\subsection{Vector-based GIS approach}
In GIS literature, many studies have discussed which data format—vector (a view
based on discrete objects) or raster (a view based on continuous fields)—leads
to a better representation of our world.
For example, \citet{egenhofer1987} proposed an object-oriented data model to
address the deficiencies of storing, manipulating, and querying spatial data in
conventional relational database management systems (RDBMS).
However, \citet{Goodchild1989} argued that the object view is a continuation
of a tradition inherited from \textit{Cartography}, and the field
representation is more realistic and accurate.
\citet{Bian2007} went a step further to generalize all environmental
phenomena into three categories and discussed each category's applicability of
adopting the object representation.
Admittedly, vector and raster representations operate on two distinct sets of
logic, but they both stem from and partially exhibit human perceptions of the
world. 
Instead of viewing them as competing or conflicting, we probably should deem
vector and raster as complementary representations of the real world like
\citet{Couclelis1992} suggested \enquote{people manipulate objects, but
cultivate fields}.
Today, we find the two representations and their corresponding analytical
methods co-exist in harmony in GIS applications, in spite of the attempts to
develop a ``general theory" of geographic representation
\citep{Liu2008,Goodchild2007,Winter2000}.

Vector-based GIS routines were implemented in PyLUSAT primarily because it fits
well conceptually with the object-oriented nature of Python, as the same choice
made by other land-use modeling programs
\citep{Barreira-Gonzalez2015,Bolte2007}.
Furthermore, such approach has two additional advantages over the raster
representation.
First, vector-based GIS tools bypass the \textit{modifiable areal unit problem}
(MAUP) since measurements or statistics are not derived from a raster grid
whose cell size is arbitrarily determined \citep{Jelinski1996}, but directly
from individual objects.
Secondly, a vector-based land-use model is more politically relevant since
objects, e.g., property parcels, reflect and honor the land ownership.
Just as \citet[p.~67]{Couclelis1992} argued that \enquote{it is at this lowest
level of real estate …, that we find the cultural grounding of the notion of
space as objects}. 
Thus, land-use decisions made on individual vector units are relatively more
feasible and realistic compared with the raster representation.

\subsection{Geospatial development with open-source Python tools}
Both \textbf{R} and \textbf{Python} are widely used in geospatial analysis, but
when it comes to development, Python is far more general-purpose and
versatile, which make it the true Swiss Army Knife of the two programming
languages.
There is an abundant resource of open-source geospatial packages in Python
\citep{carreira2016geospatial}.
For I/O-related tasks, \verb|Fiona| \citep{Gillies2011}, an API of the OpenGIS
Simple Features Reference Implementation (OGR), is capable of handling a
variety of forms of vector data from local \textit{Shapefiles} to data on
stored on a PostGIS/PostgreSQL database.
\verb|Shapely| \citep{Gillies2007}, a Python API of the Geometry Engine Open
Source (GEOS), provides methods related to set-theoretic analysis and
manipulation of planar features.
\verb|GeoPandas| \citep{Jordahl2019} is arguably the most powerful open-source
geospatial package when it comes to vector GIS, in that it merges the
functionalities of the two packages above, plus the data structures of
\verb|pandas| \citep{McKinney2010}, a fast, flexible, high-level building block
for data analysis in Python.
For raster data inputs, \verb|Rasterio| \citep{Gillies2013} based on the
Geospatial Data Abstraction Library (GDAL) can deal with I/O related tasks as
well as converting from (or to) vector data.

\section{PyLUSAT Functions for Suitability Analysis}\label{methodlogy}

Functions in PyLUSAT adhere to vector-based GIS routines, i.e., taking
vector input and generating vector output after performing one or more
geospatial operations.
In most cases, a PyLUSAT function takes a collection of polygons representing
land units as input, in which a single uniform land-use decision can be made. 
For example, these polygons can be property parcel data or an output of a 
series of GIS operations using multiple datasets, such as the
\textit{Integrated Decision Units} (IDUs) \citep{Bolte2007,Wu2015}.

The classic suitability analysis framework contains three steps: 
(1) evaluate land units based on identified criteria, 
(2) transform the measurements to a uniform suitability scale, 
and (3) combine the results to generate a single suitability score for each 
land unit \citep{Steiner2000, Marull2007}.
Correspondingly, functions of PyLUSAT fall into three categories—geospatial 
functions, transformation functions, and aggregation functions, which together 
can fully implement the framework. 
The rest of this section explains in greater detail how the three categories
of PyLUSAT's functions operate.

\subsection{Geospatial Functions}
The suitability of a parcel of land for a given land use depends greatly on its
spatial relationship with the amenities, institutions, service providers, and
natural features in the area.
Therefore, the geospatial functions are crucial to understanding land-use
suitability. 
PyLUSAT provides a variety of functions to evaluate spatial relationships 
among vector objects, such as calculating distance and density, examining 
topological predicates \citep{Strobl2008}, and interpolation.

\subsubsection{Nearest Neighbor Search}\label{nn}
\begin{figure}[!ht]
    \includegraphics[width=3.2in]{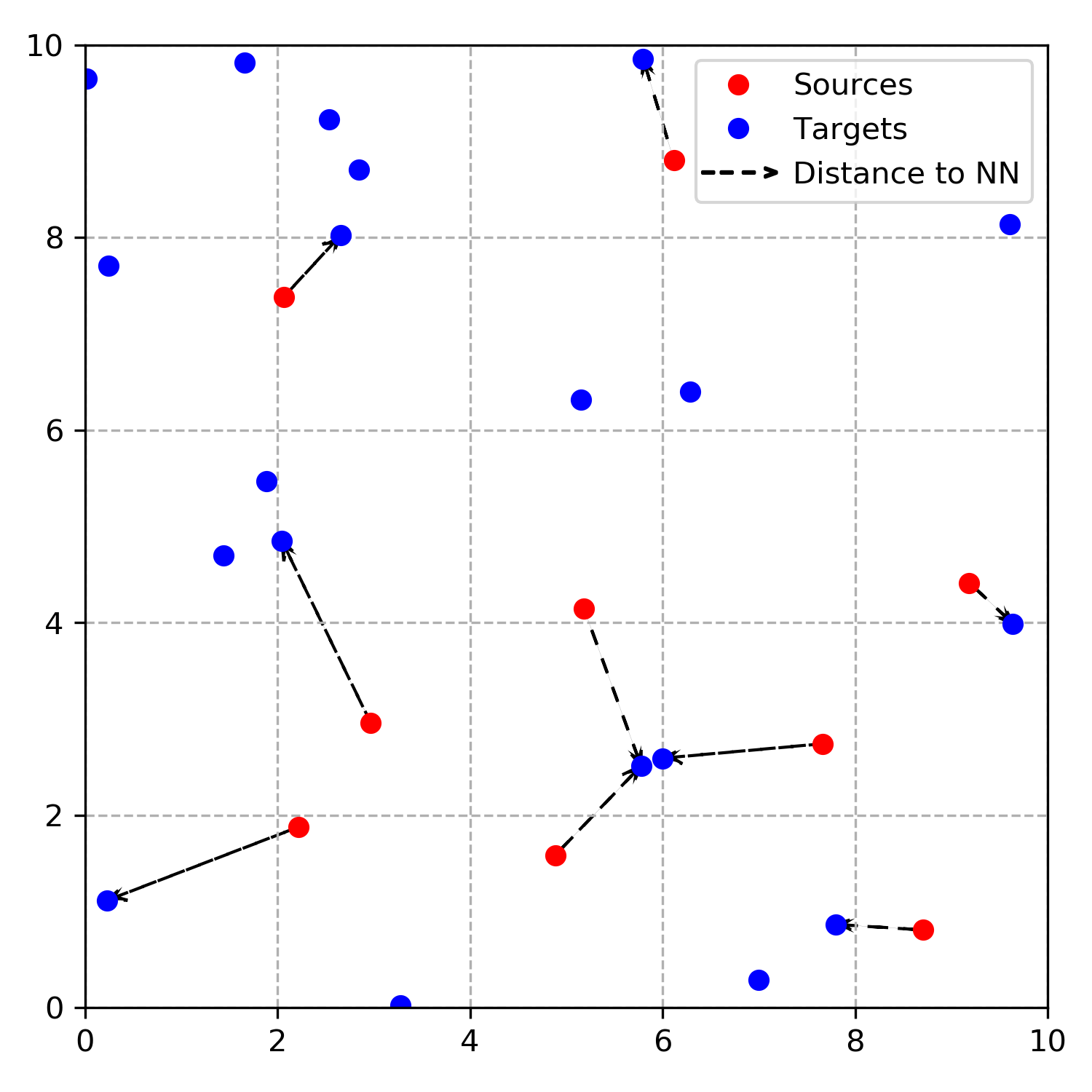}
    \centering
    \caption{The Nearest Neighbor, in terms of Euclidean distance, of each 
    source in the target set.}
    \label{fig:dist_to_pnt}
\end{figure}

\newcommand{\vect}[1]{\boldsymbol{#1}}
Distance, a direct measurement of proximity, is one of the most fundamental 
factors determining land-use suitability.
For example, due to agglomeration effects, land parcels in the vicinity of a 
central business district (CBD) \citep{Ottaviano2004} or urban sub-centers 
\citep{Yang2019} are highly suitable for commercial uses, whereas residential 
uses often favor parcels remote from nuisances, such as quarries or poultry 
farms.
In these examples, land parcels/units can be conceived as a source set paired 
with a second set, we call \textit{targets}, where distance from a source 
to the nearest target affects suitability.
Thus, calculating distance amounts to a search for the Nearest Neighbor (NN).
To generalize, given a set of $m$ sources, 
i.e., $S=\{\mathbf{s_1},\mathbf{s_2},...,\mathbf{s_m}\}$, and a set of $n$ 
targets, i.e., $T=\{\mathbf{t_1},\mathbf{t_2},...,\mathbf{t_n}\}$, 
the distance between $\mathbf{s_i}$ and $\mathbf{t_j}$, the corresponding NN 
of $\mathbf{s_i}$ in $T$, determines $\mathbf{s_i}$'s suitability, 
from a proximity standpoint.
We use $d_{e}(\mathbf{s_i})$ and $d_{m}(\mathbf{s_i})$ for the 
\textit{Euclidean} and \textit{Manhattan} distances, respectively, as defined 
by the following equations.
\begin{equation} \label{eq:1}
    d_{e}(\mathbf{s_i})=\min_j\left \|\mathbf{s_i}-\mathbf{t_j} \right \|_2,
    \mathrm{for}\:j=1,2,...,n
\end{equation}
\begin{equation} \label{eq:2}
    d_{m}(\mathbf{s_i})=\min_j\left \|\mathbf{s_i}-\mathbf{t_j} \right \|_1,
    \mathrm{for}\:j=1,2,...,n
\end{equation}
, where $\mathbf{s_i}=(s_{i1},s_{i2})'$ and $\mathbf{t_j}=(t_{j1},t_{j2})'$ 
are in $\mathbb{R}^2$; 
$\left\|\right\|_2$ denotes the Euclidean (or ${l_2}$) norm, 
i.e., $\left\|\mathbf{s_i}-\mathbf{t_j}\right\|_2=
[(s_{i1}-t_{j1})^2+(s_{i2}-t_{j2})^2]^{1/2}$; 
and $\left\|\right\|_1$ denotes the Absolute-value (or ${l_1}$) norm, 
i.e., $\left\|\mathbf{s_i}-\mathbf{t_j}\right\|_1=
(|s_{i1}-t_{j1}|+|s_{i2}-t_{j2}|)$.
Note that, here, we use a single representative point, usually a centroid, to 
represent a land unit.
Figure \ref{fig:dist_to_pnt} provides an illustration of the NN search based 
on the Euclidean distance.

Evidently, the computational complexity for this process will be 
$\mathcal{O}(mn)$, if a brute-force search was conducted 
\citep{Xiao2016}.
However, it would be overly expensive when the magnitude of $mn$ is large.
The function \verb|pylusat.distance.to_point| uses 
\verb|scipy.spatial.ckdtree| which implements a \textit{sliding midpoint} 
method to construct \textit{KDtree} (K-dimensional tree) objects to search for 
the NN more efficiently \citep{2020SciPy-NMeth}.
Like regular KDtree construction, the sliding midpoint method attempts to 
split the data at the median (midpoint) on each axis first, but the plane will 
then slide to the closest point if a trivial (all points on one side of the 
plane) split occurs \citep{DBLP:conf/dimacs/Maneewongvatana99}. 
Such a process results in a KDtree whose height need not to be 
$\mathcal{O}(log(n))$, the height of a regular KDtree, which in turn makes the 
construction of KDtree less computationally expensive. 
\citet{DBLP:conf/dimacs/Maneewongvatana99} have shown that this implementation 
offers a better performance in NN search, especially when data are clustered 
along one axis of $\mathbb{R}^2$.

\subsubsection{Affine Transformation}
When the target set is comprised of line features, an efficient approach to 
compute distances is to transform the line features from individual vector 
shapes to a raster grid. 
The function \verb|pylusat.distance.to_line| implements this approach by using 
the \textit{Affine transformation} defined by the following equation.
\begin{equation}\label{eq:3}
    \begin{bmatrix}
        v_x\\
        v_y\\
        1
    \end{bmatrix}
    =
    \begin{bmatrix}
        c & 0 &\aug& l\\
        0 & -c &\aug& t\\
        0 & 0 &\aug& 1
    \end{bmatrix}
    \begin{bmatrix}
        r_x\\
        r_y\\
        1
    \end{bmatrix}
\end{equation}
, where $c$ is the cell size used to rasterize the 2-D plane shaped by the 
extent of the line dataset; $v_x$ and $v_y$ are the $(x,y)$ coordinates of the 
vertices of the line features; $l$ and $t$ are the left and top bound of the 
2-D plane; and, finally, $r_x$ and $r_y$ represents the row and column number 
of the cell located on the transformed raster grid. 
The affine transformation matrix (ATM), i.e., the augmented matrix in the 
equation, performs a linear transformation (a scaling and a rotation) followed 
by a translation (shifting the origin), which preserves the relative spatial 
relationship among the line features \citep{Wheaton2012}. 

\begin{figure}[!ht]
    \includegraphics[width=5.4in]{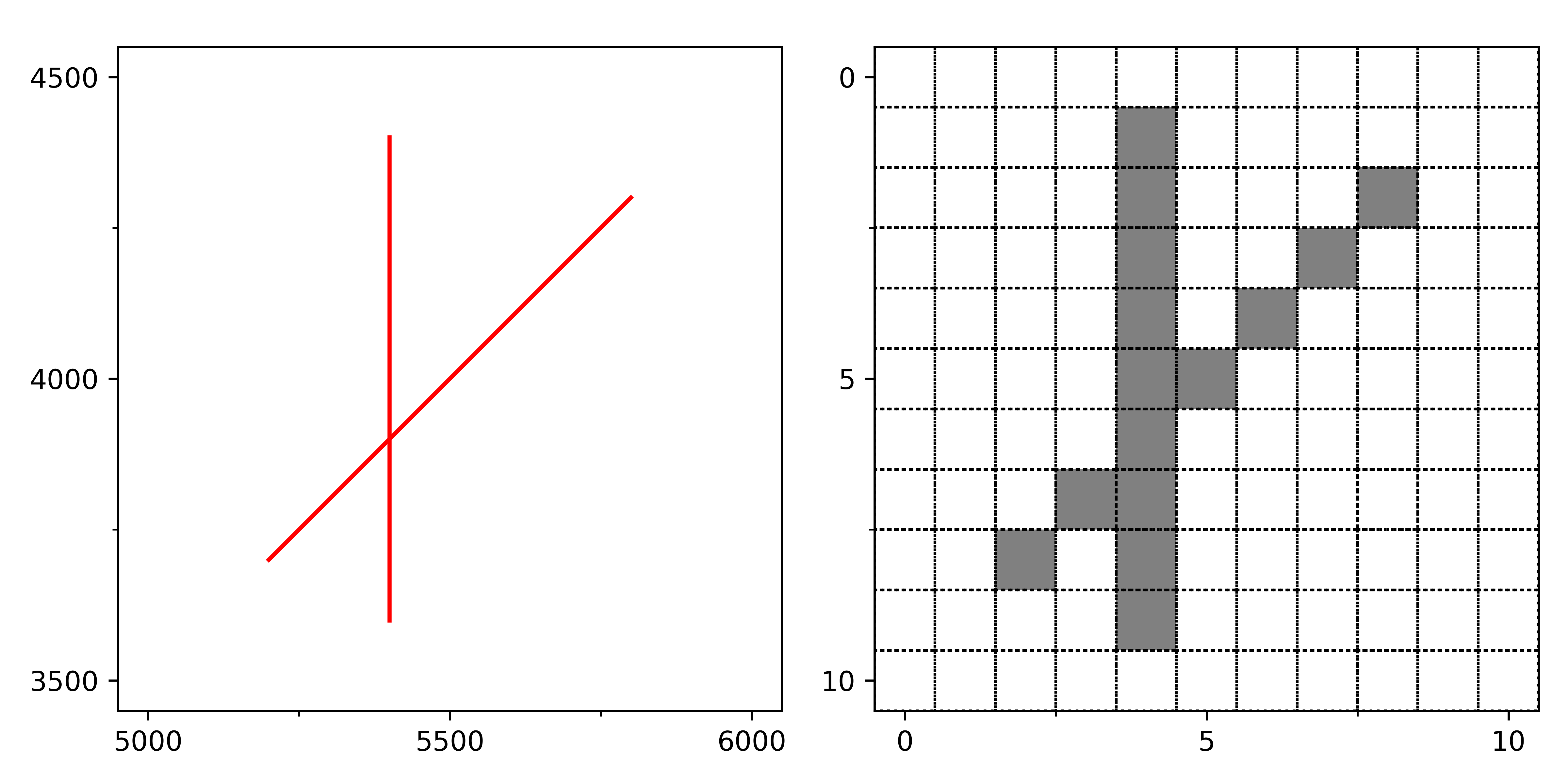}
    \centering
    \caption{An Affine Transformation that converts lines features (left) to a
    raster grid (right).}
    \label{fig:affine}
\end{figure}

After applying an Affine transformation, the problem of calculating distance 
to line features reverts to a search for NN since lines are pixelated into a 
definite amount of cells, wherever any part of any line exists. 
Note that the precision of the calculated distances will depend on $c$, the 
cell size of the converted raster grid. 
Figure \ref{fig:affine} shows an example of the transformation whose ATM's 
entries are $c=100$, $l=5000$, and $t=4500$.

\subsubsection{Density and Zonal Statistics}
Density is an important instrument in suitability analysis since it directly,
from a 2-D perspective, measures the intensity of land-use related 
activities/phenomena on a landscape.
For example, the density of single-family dwelling units in a given region
reveals characteristics of the neighborhood, and the density of a
road network reflects the level of accessibility it provides to vehicles.
Because of its vector-based characteristic, PyLUSAT's \verb|density| module
calculates density within a user-defined set of input zones (polygons).
Using an analogy of source and targets in the distance definition, the density
of targets in a given source zone $i$ is defined as 
$\rho_{i} = ({\textstyle\sum}_{j=1}^{m}{t_j}{v_j})/{A_i}$, where
$t_{j}=1$, if target $j$ is within the area of $i$ ($A_i$); otherwise
$t_{j}=0$, and $v_j$—the value corresponding to target $j$—equals to $1$, 
if not specified.

\begin{figure}[!ht]
    \includegraphics[width=4in]{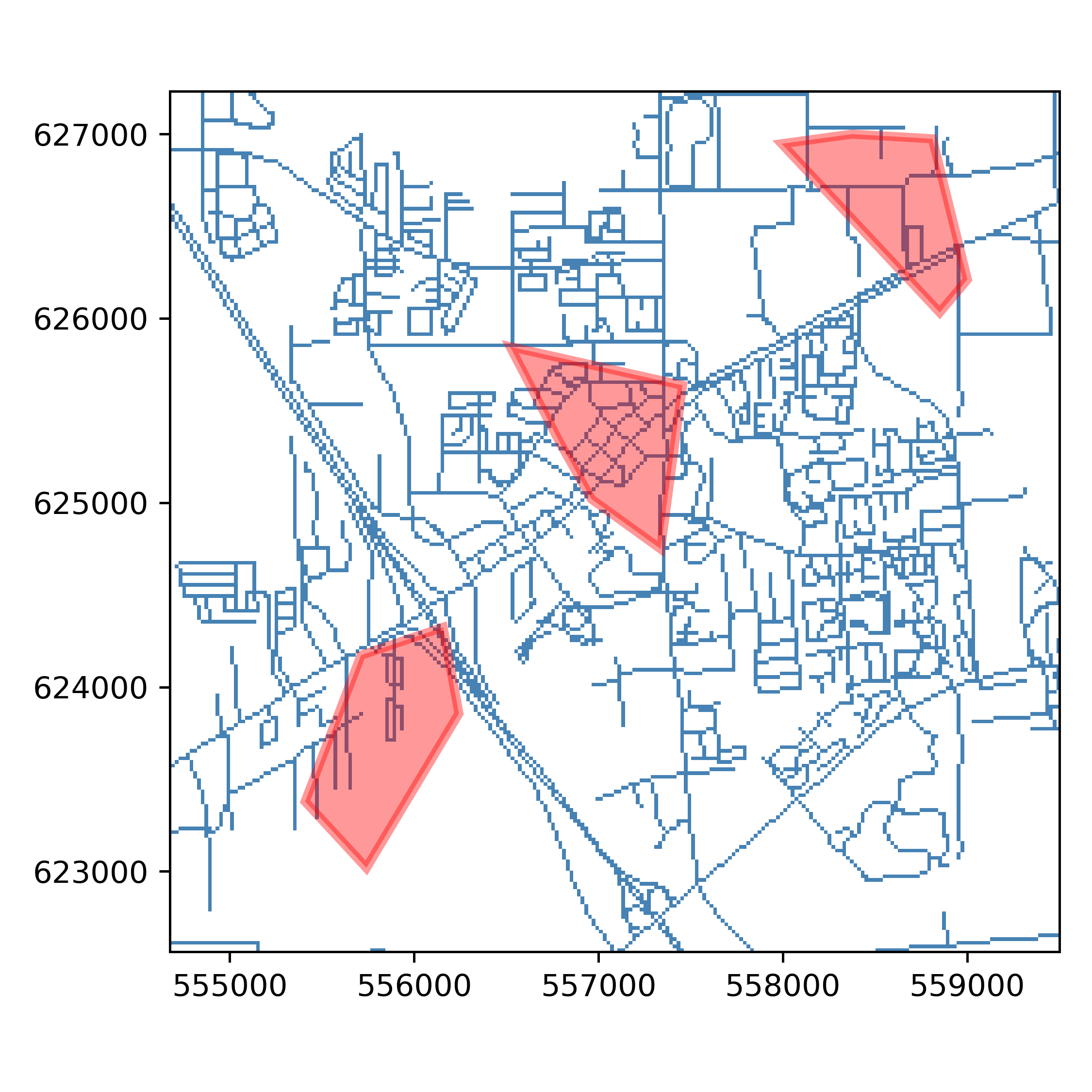}
    \centering
    \caption{Density of line features within input zones (after Affine
    Transformation applied).}
    \label{fig:line_density}
\end{figure}

To evaluate spatial containment of point targets, we used the function
\verb|geopandas.sjoin|, which supports three types of topological predicates:
\textit{intersect}, \textit{contain}, and \textit{within}. 
When targets are line features, PyLUSAT will first apply an Affine
Transformation to convert them to a raster grid, and then use
\verb|rasterstats.zonal_stats| to calculate the total amount of cells within
each source zone. 
Figure \ref{fig:line_density} illustrates the case of calculating line density.
\subsubsection{Interpolation}
Interpolation methods are used to estimate values of unknown data points using
the known ones, of which the \textit{Inverse Distance Weighted} (IDW) 
interpolation is widely applied in GIS-based suitability analysis
\citep{Tercan2020, Varatharajan2018, Yao2013}.
Borrowed from mathematics, IDW, as a GIS technique, is one of the classic
realizations of the so-called \textit{first law of geography}:
\enquote{everything is related to everything else, but near things are more
related than distant things} \citep[p.~236]{Tobler1970}.
The most commonly adopted version of IDW is Shepard’s \citeyearpar{Shepard1968}
method defined below.
\begin{equation}\label{eq:4}
    f(P)=\begin{cases}
      \dfrac{\sum_{i=1}^{N}{(d_i)}^{-u}z_i}{\sum_{i=1}^{N}{(d_i)}^{-u}} 
      & \text{if $d_i\neq0$ for all $i=1,2,...,N$}.\\
      \phantom{---}z_i & \text{if $d_i=0$ for some $i=1,2,...,N$}.
    \end{cases}
  \end{equation}
, where $P$ is the point of interest, $z_i$ and $d_i$ are, respectively, the
value at the $i$-th known data point and its Euclidean distance to $P$, and
$u$ is a predefined positive number also known as the \textit{power parameter}.
Note that, the negative sign before $u$ makes the $i$-th point's weight 
inversely proportional to its distance to $P$ in the estimation, hence how the
method got its name.
Also, as distance increases, the known values' influences on the estimation
of $P$ declines faster as $u$ gets bigger, which shifts the algorithm from a
global model to a local model.
The process of determining the best value of $u$ is relatively deterministic.
PyLUSAT provides a function, \verb|pylusat.interpolate.idw_cv|, which allows
users to pick a proper value for $u$ through \textit{Cross Validation}. 

\subsection{Transformation Function}
The goal of transformation functions is to translate measurements based on
various criteria into a standardized ``suitability" scale, which is arbitrarily
chosen and then consistently applied throughout the analysis, for example, a
scale of 1 to 9 (the lowest to highest suitability) adopted by the Land Use
Conflict Identification Strategy (LUCIS) \citep{Carr2007}.
In general, there are three mechanisms to define a transformation, which are
by (a) unique categories, (b) range of classes, and (c) continuous functions.
The first mechanism typically deals with nominal and ordinal data, in which
individual values are associated with different degrees of suitability.
The second method handles interval or ratio data and is more flexible, in that
values fall into a certain range represent the same degree of suitability.
PyLUSAT's \verb|rescale.reclassify| function allows users to do both
transformations by leveraging the \verb|pandas.DataFrame| object's highly
efficient indexing/slicing capability.

Besides pre-defined ranges, ranges can be derived from the data as well.
Jenks \citeyearpar{Jenks1977} developed the \textit{natural breaks} algorithm,
originally as a choropleth mapping technique, which was widely employed in
land-use suitability analysis \citep{Abdullahi2015,Berry2015,Owusu2017}.
Natural breaks seek to simultaneously minimize the differences within classes
and maximize the differences between classes.
The total variance in the data, also known as the squared deviation from array
mean (SDAM), is defined as $\sum_{i=1}^n(x_i-\bar{x})^2$, on the other hand, 
the within-class difference is captured by the squared deviation from the class
mean (SDCM), that is $\sum_{j=1}^{k}\sum_{i=1}^{n_j}(x_i-\bar{x}_j)^2$,
where $n_j$ is the number of elements in the $j$-th class, and $k$ is a
pre-defined number of classes. 
The algorithm iterates through all possible breaks and computes the Goodness of
Variance Fit (GVF), i.e., $\frac{SDAM-SDCM}{SDAM}$.
The minimum GVF obtained corresponds to the so-called ``optimal" range of
classes.

As for transforming by continuous functions, PyLUSAT currently supports only a
\textit{linear} transformation, but we intend to release a complete list of
functions in the next major update of the package.
Although the linear transformation, also known as the 
\textit{min-max feature scaling}, is relatively simple, it has been used as a
feature engineering technique in many machine learning applications
\citep{Tang2018}.
Besides the intuitiveness, another advantage that might make the method
appealing in suitability analysis is that it preserves the distribution of
the original variable after the transformation \citep{Cao2015}.
Without loss of generality, the following equation defines a linear
transformation of a variable $X$ from its original scale $[x_{min},x_{max}]$ to
an arbitrary scale $[a,b]$.
\begin{equation}\label{eq:5}
    x_i^\prime=\begin{cases}
      a+\dfrac{x_i-x_{min}}{x_{max}-x_{min}}(b-a) 
      & \text{for $i\in\{1,2,...n\}$ (regular order)}.\\[10pt]
      b-\dfrac{x_i-x_{min}}{x_{max}-x_{min}}(b-a) 
      & \text{for $i\in\{1,2,...n\}$ (inverse order)}.
    \end{cases}
\end{equation}
, where $n$ is the total number of observations.
Note that, both cases in equation \ref{eq:5} are relevant to measuring land-use
suitability.
As with the two examples in Section \ref{nn}, a relatively large value of
the same measurement, i.e., distance, may be valued as either pros or cons,
depending on the suitability criteria.

\subsection{Aggregation Function}
In GIS-based suitability analysis, transformed measurements of various criteria
are combined to make a land-use decision, which commonly is done by assigning
weights to individual criteria, based on expert knowledge, and then summing the
results \citep{Kalogirou2002}.
PyLUSAT provides a utility function \verb|pylusat.util.weighted_sum| for such
operation.
However, professionals or stakeholders (agents) often find themselves in a
situation where a consensus on the weighting cannot be reached.
When this happens, a Multi-Criteria Decision Making (MCDM) technique is
helpful.
To address MCDM problems, PyLUSAT offers a function, \verb|pylusat.utils.ahp|,
to implement the Analytic Hierarchy Process (AHP) developed by
\citet{Saaty1990}.

Based on the premise that people are good at comparing two (but no more
than two) items, AHP converts a decision of multiple criteria into a series of
pair-wise comparisons, with the result quantified using a scale from $1$ to
$9$. 
If item \textit{A} is equally important to item \textit{B}, the result is $1$.
And, if item \textit{A} is extremely important than item \textit{B}, the result
is $9$.
The integers in between correspond to different levels of pair-wise importance
comparisons.
Moreover, the reciprocals of these values are used if one swaps the comparates,
i.e., if \textit{A} to \textit{B} is $5$ ($w^{}_{A}/w^{}_{B}=5$), then
$w^{}_{B}/w^{}_{A}=1/5$.
According to this setup, AHP first creates a reciprocal matrix using results of
the pair-wise comparisons. 
Then, it solves the eigenvalue equation, i.e., $A\bm{v}=\lambda\bm{v}$, and
retains the primary eigenvalue (the largest one among all eigenvalues) and the
corresponding primary eigenvector.
Finally, it normalize the primary eigenvector (dividing individual elements of
the vector by their sum), to obtain the priority vector.
Each element of the priority vector represents the weight of an initial
criterion involved in the analysis, which reflects its relative importance
in the final decision.

AHP also involves a mechanism, \textit{Consistency Ratio} (CR), to validate
whether the decisions of the pair-wise comparisons are consistent, e.g., given
$w^{}_{A}/w^{}_{B}=7$ and $w^{}_{B}/w^{}_{C}=3$, then if
$w^{}_{A}/w^{}_{C}=1/5$, we call it an ``inconsistency" in the comparisons.
In addition to the \verb|pylusat.utils.ahp| function, PyLUSAT also provides a
\verb|pylusat.utils.random_ahp| function to generate random AHP weights that
follows the rule of thumb, that is CR is less than $0.1$.

\section{Validation and Evaluation}
\begin{figure}[!ht]
    \includegraphics[width=5.4in,scale=0.2]{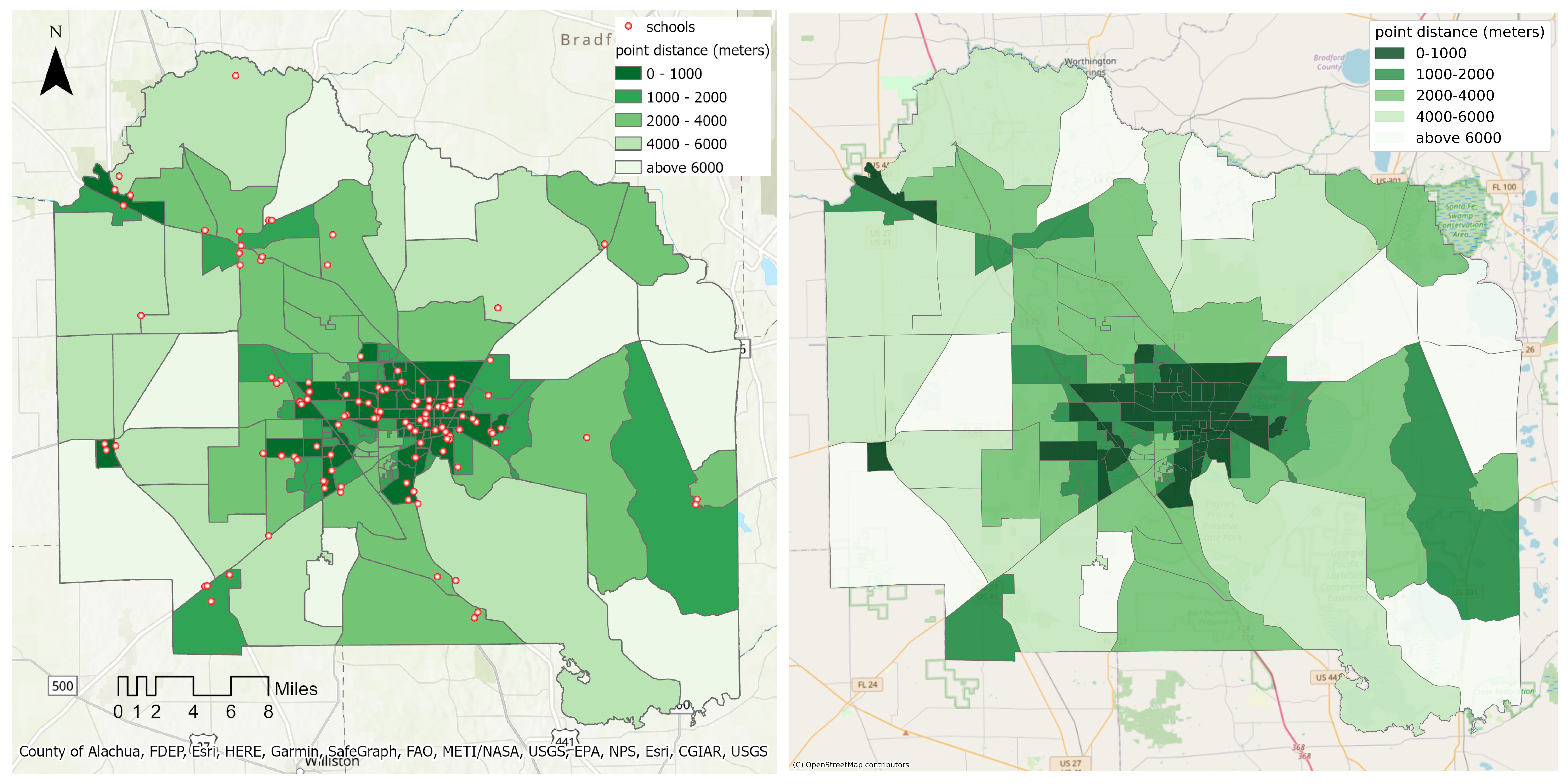}
    \centering
    \caption{Maps of measuring point distance. Left: by ArcGIS. Right: by
    PyLUSAT.}
    \label{fig:pnt_dist_compare}
\end{figure}

Figure \ref{fig:pnt_dist_compare} shows two choropleth maps side-by-side, where
the left one presents the result of measuring point distances between schools
and centroids of census block groups (CBG) of Alachua County, Florida; and the
right one shows the same phenomenon with the same color scheme but measured by
PyLUSAT.
The datasets used to create the two maps, the left by a \textit{layout} of
ArcMap and the right by the plotting function of \verb|GeoPandas| and
\verb|Contextily| (for basemap tiles), are included in the GitHub
repository mentioned in the first section.
As the figure shows, from a cartographic perspective, the two
results are identical.
However, to validate tools in PyLUSAT, especially the geospatial functions, 
more rigorously, we compared outputs of five \textit{geoprocessing tools} in
ArcMap 10.4 with the outputs of corresponding functions in PyLUSAT by
conducting a series of (two-tailed) paired-sample \textit{t}-tests.
The null hypotheses of these tests are identical, which is there exists no
statistically significant difference between outputs from PyLUSAT functions and
their ArcGIS counterparts.
Table \ref{table:t_tests} lists (by function) the degree of freedoms (df),
observed \textit{t} statistics, and \textit{p}-values of these \textit{t}
tests.
\begin{table}[!ht]
    \centering
    \begin{tabular}{llrr}
    \hline
    Test function     & df  & \textit{t} statistic & \textit{p}-value  \\
    \hline
    Distance to point & 154 & 1.0528               & 0.2941   \\
    Distance to line  & 154 & 0.3613               & 0.7184   \\
    IDW               & 154 & 0.5889               & 0.5568   \\
    Density of point  & 154 & -0.6863              & 0.4936   \\
    Density of line   & 154 & 1.1656               & 0.2447   \\
    \hline  
    \end{tabular}
    \caption{Paired-sample \textit{t}-tests between results from PyLUSAT and
    ArcGIS.}
    \label{table:t_tests}
\end{table}
In these \textit{t} tests, the observations are different quantities measured
against the 155 CBGs in Alachua County, hence 154 df.
We used school and road network datasets in the county for point and line
features respectively which, again, are included in the GitHub repository.
For IDW, we used the Digital Elevation Model (DEM) as the value raster grid.
As indicated by the \textit{p}-values, none of the tests can reject the null
hypothesis, which suggests that we can trust the results of PyLUSAT's
geospatial functions with confidence.

\begin{figure}[!ht]
    \includegraphics[width=5.4in,scale=0.2]{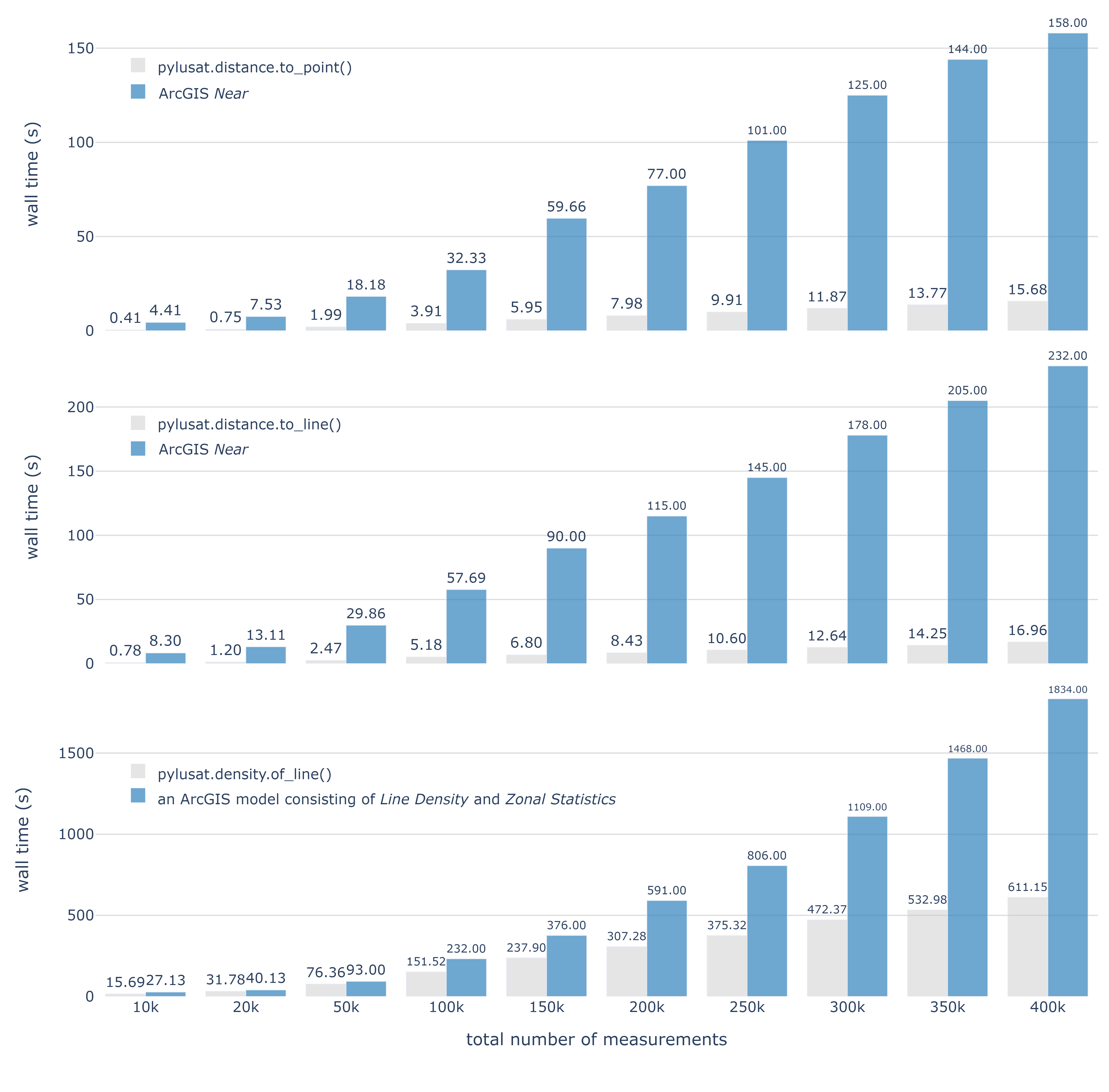}
    \centering
    \caption{Running times for PyLUSAT functions compared with their ArcGIS
    counterparts.}
    \label{fig:eval_performance}
\end{figure}

PyLUSAT is developed with computing speed in mind as well.
In contrast to conducting suitability analysis using GIS applications,
computational efficiency is mainly gained from two sources: (a) the
implementation of \textit{NumPy's} vectorized operation in PyLUSAT and (b) the
I/O wait time saved from reading/writing intermediate files \citep{Harris2020}.
The latter is non-negligible in that GIS applications need to store
intermediate files \textit{on disk}, whereas PyLUSAT keeps the study units
(e.g., land parcels) \textit{in memory} throughout the entire process of
suitability analysis.
Figure \ref{fig:eval_performance} shows three time cost (wall time measured in
seconds) comparisons between PyLUSAT functions and their counterparts in 
\textit{ArcMap 10.4}.
Note that, \verb|pylusat.density.of_line()| calculates line density in each
input polygon or in a user-defined radius around each polygon's centroid.
Since the function is different from ArcMap's \textit{Line Density} tool, which
only produces a raster grid with density values, a ModelBuilder model
consisting of \textit{Line Density} and \textit{Zonal Statistics} was used in
the third comparison.
As shown in the figure, PyLUSAT functions take less time to run in all three
cases, and such effect become more significant as the total number of
measurements increases.

The improved computational efficiency by individual PyLUSAT functions could
significantly reduce the total time cost of conducting a suitability analysis.
Moreover, this effect can be further amplified by HPC.
Since PyLUSAT is cross-platform, it can be installed on a computing cluster
(usually running on Linux-based OS) with minimal effort.
Thus, it enables urban planners and researchers to rapidly simulate future
land-use scenarios to evaluate both the intended and unintended consequences of
specific land-use policies under the framework of suitability analysis. 
This is a main goal of the development of PyLUSAT.
Chen \citeyearpar{Chen2019} conducted a feasibility study, in which PyLUSAT is
used to port the LUCIS model to \textit{HiPerGator}, a supercomputer at the
University of Florida \citep{Carr2007}.
In this study, ninety-six cores were used to simulate 120 alternative land-use
scenarios in Orange County, Florida.
The entire process took only slightly over five minutes.

\section{Conclusion}
Open-source software dedicated to GIS-based land use suitability analysis is
rarely found in relevant literature.
In this paper, we present a Python package—PyLUSAT—representing a promising
candidate to help fill in this absence.
As an alternative solution to existing GIS applications, PyLUSAT facilitates
the customization and automation of suitability analysis while maintaining the
process highly scalable and reproducible.
The performance of PyLUSAT's functions were evaluated from both accuracy and
efficiency perspectives.
Five \textit{geospatial functions} in PyLUSAT were selected and conducted
paired-sample \textit{t} tests between the outputs from these five functions
and outputs from their counterparts in \textit{ArcMap 10.4}.
Results of these tests showed that there are no statistically significant
difference between the two sets of outputs.
Additionally, we benchmarked the time costs (wall time) of three PyLUSAT's
geospatial functions and their corresponding tools in ArcMap.
Results showed that PyLUSAT functions are noticeably faster.

PyLUSAT has been made available on the Python Package Index (PyPI) and also on
GitHub at https://github.com/chjch/pylusat.
It offers various tools (functions), allowing the package to handle tasks
entailed by GIS-based land-use suitability analysis.
PyLUSAT can be used not only on a personal computer running on either Windows,
Linux, or MacOS, but also on a supercomputer to take advantage of HPC.
In addition, PyLUSAT is highly extensible.
For example, it is currently being used to develop fourteen QGIS
\textit{Processing Algorithms} to support sustainable land management (SLM) in
Ghana \citep{GALUPteam2021}.
Finally, methods and tools introduced in Section 3 of this paper can be used
by developers in the FOSS community who are interested in developing
geospatial packages and applications.





 
\printbibliography
\end{document}